\documentclass[12pt]{article}

\usepackage{amsmath}
\usepackage{amsfonts}
\usepackage{amssymb}

\newcommand{\be}{\begin{equation}}
\newcommand{\ee}{\end{equation}}
\newcommand{\ba}{\begin{eqnarray}}
\newcommand{\ea}{\end{eqnarray}}

\begin{document}

\begin{center}
{\large\bf ANALYTICAL SOLUTION OF TWO-DIMENSIONAL SCARF~II MODEL BY MEANS OF SUSY METHODS}\\

\vspace{0.3cm} {\large \bf M.V. Iof\/fe$^{1}$\footnote{E-mail: m.ioffe@pobox.spbu.ru},
E.V. Krupitskaya$^{1}$\footnote{E-mail: e.v.krup@yandex.ru},
D.N. Nishnianidze$^{1,2}\footnote{E-mail: cutaisi@yahoo.com}$
}\\
\vspace{0.2cm}
$^1$Saint-Petersburg State University,
198504 St.-Petersburg, Russia\\
$^2$Akaki Tsereteli State University, 4600 Kutaisi, Republic of Georgia\\
\end{center}

\hspace*{0.5in}

\begin{minipage}{6.0in}
{\small New two-dimensional quantum model - the generalization of the Scarf II -
is completely solved analytically for the integer values of parameter. This model being not
amenable to conventional procedure of separation of variables is solved by recently proposed
method of supersymmetrical separation. The latter is based on two constituents of SUSY Quantum Mechanics:
the intertwining relations with second order supercharges and the property of shape invariance. As a result,
all energies of bound states were found, and the analytical expressions for corresponding wave functions were
obtained.
\\
\vspace*{0.1cm} PACS numbers: 03.65.-w, 03.65.Fd, 11.30.Pb }
\end{minipage}

\vspace*{0.4cm}

\setcounter{footnote}{0} \setcounter{equation}{0}

\section{Introduction.}

The supersymmetrical methods originating from the Elementary Particle Theory gave an essential impulse
in development of non-relativistic Quantum Mechanics \cite{junker} - \cite{krive} (and references therein) starting from the famous paper of E.Witten \cite{witten}.
Actually, this approach reproduced - in a modern form - two very old methods: Darboux transformation for Sturm-Liouville
equation in Mathematical Physics \cite{darboux}
and Factorization Method for Schr\"odinger equation in Quantum Mechanics \cite{schredinger}, \cite{infeld}.
The modern frameworks provided new insight at the abilities of analytical methods in Quantum Mechanics. In particular,
a new notion of shape invariance property was naturally introduced \cite{gendenstein} in this approach. This property allowed to reproduce all known exactly solvable one-dimensional models \cite{bagchi}, \cite{dabrowska}. In turn, by means of supersymmetrical intertwining relations these models produce a class of new exactly solvable models, which are isospectral (or almost isospectral) to the original ones \cite{cooper}, \cite{bagchi}.
Additional opportunities were opened up from one-dimensional polynomial SUSY Quantum Mechanics with supercharges of higher order in derivatives, proposed in \cite{asv}, \cite{acdi} and developed essentially by many authors in \cite{samsonov} - \cite{quesne1}. Different solvable models with given properties of spectra (quantum design) were built (see the review paper \cite{andrianov}).

The success of supersymmetrical approach in {\it multidimensional} Quantum Mechanics must be discussed separately. It includes
both the achievements of direct generalizations \cite{abei} - \cite{scattering} of Witten's formulation with first order supercharges, and of polynomial SUSY Quantum Mechanics with supercharges of second order in momenta \cite{david}, \cite{three-dim}, which generalized the one-dimensional polynomial SUSY Quantum Mechanics mentioned above. The direct multidimensional generalization provided the investigation of spectra of some matrix quantum problems \cite{abei} - \cite{scattering}, while polynomial one allowed to exclude the matrix components from the superHamiltonian \cite{david}. The essential results were obtained \cite{david} - \cite{ioffe2} by polynomial SUSY approach in analysis of {\it two-dimensional} quantum models which {\it are not amenable} to the standard separation of variables. Until now, the latter procedure \cite{eisenhart}, which reduces the problem to several problems of lower dimensionality, was the only practical method of solving the multidimensional models. Recently, three new methods for the problem were proposed in the framework of supersymmetrical method \cite{new}, \cite{innn}, \cite{ioffe1}, \cite{ioffe2}. All three can be considered as different variants of {\it supersymmetrical separation of variables}. They explore the supersymmetrical intertwining relations, which provide that the corresponding quantum systems obey the symmetry of fourth order in momenta instead of second order in the case of standard separation of variables.

Two of the SUSY methods above \cite{new} - \cite{poschl1}, \cite{ioffe1} lead to {\it partial (quasi-exact) solvability} of the models with only a part of spectra known, and one - to {\it complete (exact) solvability} of two-dimensional generalizations of Morse \cite{physrev} and P\"oschl-Teller \cite{poschl2} models (see also the review paper \cite{ioffe2}). The idea to consider the quasi-exactly-solvable models - the intermediate class between exactly solvable and unsolvable analytically models - was introduced on 80-ties in \cite{razavy} - \cite{kamran}. In particular, the series of papers by A.Turbiner, A.Ushveridze and M.Shifman was devoted to the elegant algebraic method of construction of one-dimensional quasi-exactly-solvable (and sometimes, of exactly solvable) quantum models. In general, this method works beyond the supersymmetry, but both approaches can be combined in one-dimensional case as in \cite{shifman3}. This approach is applicable to two-dimensional problems as well, but only in curved spaces with nontrivial metrics \cite{shifman1}. Recently, new classes of solvable two-dimensional models were built in \cite{ttw}, \cite{quesne2}, \cite{marquette}, and all these models are superintegrable and amenable to separation of variables. Thus, the first two variants of supersymmetrical separation of variables proposed in \cite{new} - \cite{poschl1}, \cite{ioffe1} provide quite different method of construction of new two-dimensional quasi-exactly-solvable models, and the third one \cite{physrev}, \cite{poschl2}, \cite{ioffe2} - of new exactly solvable models. All constructed models do not allow standard separation of variables, but by construction they are integrable with the symmetry operators of fourth order in momenta.

It will be shown in the present paper, that one more two-dimensional model - with potential, which is naturally associated with solvable one-dimensional hyperbolical version of Scarf model (Scarf II) \cite{scarf}, \cite{dabrowska} - can be solved analytically by means of supersymmetrical separation. This potential was obtained recently \cite{shape2011} among new two-dimensional models with shape invariance property \cite{gendenstein}, \cite{mallow}, \cite{new}, \cite{ioffe1}. Just this property will allow to solve the problems with the whole hierarchy of generalized Scarf II potentials. The structure of the present paper is the following. In Section 2, the two-dimensional generalized Scarf II model will be completely solved for the specific parameter value $a=-1:$ both
energy values and corresponding wave functions of all bound states will be built analytically. In Section 3, the procedure will be generalized to the models with arbitrary negative integer values of parameter $a.$ Some cumbersome calculations are given in Appendix.

\section{Exact solution of the model for $a=-1$}

\subsection{Formulation of the model}

We start from the supersymmetrical intertwining relations
\be
H^{(1)}Q^{+}=Q^{+}H^{(2)}; \qquad Q^{-}H^{(1)}=H^{(2)}Q^{-}, \label{int}
\ee
for two partner two-dimensional Hamiltonians of Schr\"odinger form
\be
H^{(i)}=-\Delta^{(2)}+V^{(i)}(\vec x);\,\, i=1,2;\quad \vec x = (x_1, x_2);\quad \Delta^{(2)}\equiv \partial_1^2+\partial_2^2;\quad
\partial_i\equiv \frac{\partial}{\partial x_i} \label{H}
\ee
with mutually conjugated supercharges $Q^{\pm}$ of second order in derivatives.
A list of particular solutions of (\ref{int}) is known \cite{david}, \cite{ioffe1}, and a part of them was
studied in detail. Depending on chosen values of parameters, the partial and/or complete solutions for spectra and bound state
wave functions of corresponding models were obtained \cite{new}, \cite{physrev}, \cite{poschl1}, \cite{poschl2}, \cite{innn}, \cite{ioffe1}, \cite{ioffe2}.

Below, a new solution of (\ref{int}) obtained in the recent paper \cite{shape2011} will be analyzed by means of supersymmetrical separation of variables. The potentials are:
\ba
V^{(1),(2)}(\vec x)&=&-2\lambda^{2}a(a\mp1)(\frac{1}{\cosh^{2}(\lambda x_{+})}-\frac{1}{\sinh^{2}(\lambda x_{-})})
-\frac{2k_{1}\sinh(2\lambda x_{1})+k_{2}}{4\cosh^{2}(2\lambda x_{1})}-\nonumber\\
&-&\frac{2k_{1}\sinh(2\lambda x_{2})+k_{2}}{4\cosh^{2}(2\lambda x_{2})},
\label{V}
\ea
and the second order supercharges are:
\ba
Q^{+}&=&(Q^{-})^{\dag}=4\partial_{+}\partial_{-}+4\lambda a\tanh (\lambda x_{+})\partial_{-}+4\lambda a\coth (\lambda x_{-})\partial_{+}+\nonumber\\
&+&4\lambda^2a^{2}\tanh (\lambda x_{+})\coth (\lambda x_{-})
+\frac{2k_{1}\sinh( 2\lambda x_{1})+k_{2}}{4\cosh^{2}( 2\lambda x_{1})}-\frac{2k_{1}\sinh( 2\lambda x_{2})+k_{2}}{4\cosh^{2}(2\lambda x_{2})}, \label{Q}
\ea
where $x_{\pm}\equiv x_{1} \pm x_{2}, \partial_{\pm}=\partial / \partial x_{\pm},$ and $\lambda , a, k_1, k_2$ are real parameters.
The potentials (\ref{V}) are not amenable to standard separation of variables, but they correspond to integrable Hamiltonians (\ref{H}) with symmetry operators of fourth order in derivatives:
\be
R^{(1)}=Q^+Q^-;\quad R^{(2)}=Q^-Q^+.
\ee

The first step of the present approach is to choose such values of parameters, that one of the Hamiltonians $H^{(2)}$ does allow standard separation of variables. Then, we have a chance to find the spectrum and wave functions of the partner Hamiltonian $H^{(1)}$ which does not allow standard separation. The expressions (\ref{V}) are just good candidates to realize this approach. Indeed, one can choose the parameter $a=-1$ to cancel the terms prohibited from separation. For simplicity, we shall also fix the parameter $\lambda =1/2.$ Thus,
\ba
H^{(1)}(\vec x) &=& -\Delta^{(2)}-\bigl(\frac{1}{\cosh^{2}(x_{+}/2)}-\frac{1}{\sinh^{2}(x_{-}/2)}\bigr) + U(x_{1}) + U(x_{2})\nonumber\\
H^{(2)}(\vec x) &=& -\Delta^{(2)} + U(x_{1})+  U(x_{2}),\nonumber
\ea
where one-dimensional potential $U$ is defined as:
\be
U(x)= -\frac{2k_{1}\sinh(x)+k_{2}}{4\cosh^{2}(x)}. \label{U}
\ee

\subsection{Solution of the model with separated variables}

The second step of the method - solution of the two-dimensional problem with Hamiltonian $H^{(2)}$ by means of separation of variables: the one-dimensional Schr\"odinger equation with potential $U(x)$ has to be solved. For the general case of $U,$
this is impossible to perform analytically. But for specific form (\ref{U}) for $U(x),$ the solution is known explicitly \cite{scarf}, \cite{dabrowska}. The Schr\"odinger equation
\begin{equation}\label{scarf}
[-\partial^2 + (B^2-A^2-A)\frac{1}{\cosh^{2}(x)}+B(2A+1)\frac{\sinh(x)}{\cosh^{2}(x)}]\eta_n (x) = \varepsilon_n\eta_n (x)
\end{equation}
has the finite discrete spectrum with energy eigenvalues:
\begin{equation}\label{epsilon}
\varepsilon_n = -(A-n)^2
\end{equation}
and wave functions:
\begin{equation}\label{eta}
\eta_n(x)=(\cosh(x))^{-A}\exp(-B\arctan (\sinh(x))) P_{n}^{(-iB-A-1/2,+iB-A-1/2)}(i\sinh(x))
\ee
In the formulas above, $A, B$ are positive parameters $A, B > 0$, and $P_{n}^{(\alpha,\beta)}$ are the $n-$th power Jacobi polynomials of their argument \cite{jacobi}. Comparing expressions (\ref{U}) and (\ref{scarf}), the positive parameters $A, B$ can be expressed in terms of coupling constants $k_1 <0, k_2:$
\begin{eqnarray}
A &=& -1/2-\frac{1}{\sqrt{2}}k_{1}\biggl(\sqrt{(k_{2}+1)^{2}+4k_{1}^{2}}-(k_{2}+1)\biggr)^{1/2}\nonumber\\
B &=& \frac{1}{2\sqrt{2}}\biggl(\sqrt{(k_{2}+1)^{2}+4k_{1}^{2}}-(k_{2}+1)\biggr)^{1/2}.\nonumber
\end{eqnarray}
The condition of normalizability of bound state wave functions (\ref{eta}) gives $n < A$ in (\ref{epsilon}), (\ref{eta}).

The spectrum of two-dimensional Hamiltonian $H^{(2)}$ is two-fold degenerated (for $n\neq m$), and its normalizable wave functions are built from $\eta (x)$ as their symmetric and antisymmetric combinations:
\begin{eqnarray}
E^{(2)}_{n,m} &=& E^{(2)}_{m,n} = \epsilon_{n}+\epsilon_{m}=-(A-n)^{2}-(A-m)^{2}; \label{energy}\\
\Psi_{n,m}^{(2)\pm}&=& \pm\Psi_{m,n}^{(2)\pm} = \eta_{n}(x_{1})\eta_{m}(x_{2})\pm\eta_{m}(x_{1})\eta_{n}(x_{2}). \label{psi2}
\end{eqnarray}

\subsection{Solution of the model with non-separable variables}

The next step is to find analytically the discrete spectrum and normalizable wave functions for the quantum problem with
the Hamiltonian $H^{(1)}(\vec x)$ with parameter $a=-1.$ The main tools for solution of this task are the supersymmetrical intertwining relations (\ref{int}), which provide the links between spectra and wave functions of partner Hamiltonians \cite{david}. In general, these Hamiltonians are isospectral,
but some properties of intertwining operators $Q^{\pm},$ such as their singularities and zero modes, are crucial at this stage.
Generally speaking, three kinds of bound states of $H^{(1)}$ may exist (items (i), (ii) and (iii) below).

(i). First of all, in this approach the wave functions of $H^{(1)}$ are obtained from (\ref{energy}), (\ref{psi2}) by means of intertwining relations (\ref{int}) in the form:
\begin{equation}\label{psi1}
\Psi^{(1) \mp}_{n,m}(\vec x) = Q^+ \Psi^{(2) \pm}_{n,m}(\vec x),\quad E^{(1)}_{n,m}=E^{(2)}_{n,m}=-(A-n)^{2}-(A-m)^{2},
\end{equation}
where the choice of $\mp-$superscript in the l.h.s. points out the change of symmetry under $x_1 \leftrightarrows x_2,$ due to antisymmetry
property of operator $Q^+.$ The intertwining operator $Q^+$ has singularity along the line $x_- = 0,$ therefore the behaviour of (\ref{psi1})
must be analyzed along this line. The direct calculations show that only the symmetric functions $\Psi^{(1) +}_{n,m}(\vec x)$ are normalizable.
Up to the end of this Section, we shall use the following notations: $\Psi^{(1) +}_{n,m}(\vec x)\equiv \Psi^{(1)}_{n,m}$ and $\Psi^{(2) -}_{n,m}(\vec x)\equiv \Psi^{(2)}_{n,m}.$ It is obvious that for $n=m$ the functions $\Psi^{(1) +}_{n,n}(\vec x)\equiv 0.$ Also, it is not excluded that some functions $\Psi^{(1)}$ vanish since the functions $\Psi^{(2) -}$ could be simultaneously the zero modes of $Q^+.$

For investigation of the latter opportunity, we shall use the indirect algebraic method which will be also very useful in the next Section. Namely, let us express the norm of wave functions (\ref{psi1}) in terms of the following matrix elements:
\be
\|\Psi_{n,m}^{(1)}\| = <\Psi_{n,m}^{(2)}\mid Q^{-}Q^{+}\mid \Psi_{n,m}^{(2)}> = r_{n,m}\|\Psi_{n,m}^{(2)}\|^{2}, \nonumber
\ee
where we used the fact that for $H^{(2)}$ the operator $R^{(2)}=Q^-Q^+$ is the symmetry operator (with the eigenvalue $r_{n,m}$), which does not change the symmetry (antisymmetry) of wave functions. The expressions for $Q^{\pm}$ are given by (\ref{Q}), and after straightforward calculations
the symmetry operator takes the form:
\begin{eqnarray}
& &R^{(2)} = \partial_{1}^{4}+\partial_{2}^{4}-2\partial_{1}^{2}\partial_{2}^{2}-2\bigl(1+U(x_{1})-U(x_{2})\bigr)\partial_{1}^{2}-\nonumber \\
&-&2\bigl(1-U(x_{1})+U(x_{2})\bigr)\partial_{2}^{2}-2\bigl(\partial_{1}U(x_{1}))\partial_{1}-2(\partial_{2}U(x_{2})\bigr)\partial_{2}-\nonumber\\
&-& 4\bigl(\partial_{-}C_{-}-C_{-}^{2}\bigr)+2C_{+}\bigl(U^{\prime}(x_{1})+U^{\prime}(x_{2})\bigr)+2C_{-}\bigl(U^{\prime}(x_{1})-U^{\prime}(x_{2})\bigr)
+\nonumber \\
  &+& 8C_{+}C_{-}\bigl(U(x_{2})-U(x_{1})\bigr)-U^{\prime\prime}(x_{1})-U^{\prime\prime}(x_{2})+U^{2}(x_{1})+U^{2}(x_{2})-2U(x_{1})U(x_{2}),\label{long}
\end{eqnarray}
where $U(x_i)$ are given by (\ref{U}), $U^{\prime}, U^{\prime\prime}$ are their derivatives, and $C_{\pm}(x_{\pm})$ are the coefficient functions of the supercharge $Q^+$ for the present model:
\begin{equation}\label{C}
C_{+}=-1/2\tanh(x_{+}/2);\quad C_{-}=-1/2\coth(x_{-}/2).
\end{equation}
Quite similarly to the procedure in \cite{physrev}, one can check that (\ref{long}) can be expressed in terms of one-dimensional Hamiltonians $h_i(x_i)\equiv -\partial_i^2 + U(x_i)$ as follows:
\be
R^{(2)}=\bigl(h_{1}(x_1)-h_{2}(x_2)\bigr)^{2}+2\bigl(h_{1}(x_1)+h_{2}(x_2)\bigr)+1. \nonumber
\ee
Since the wave functions $\Psi^{(2)}_{n,m}(\vec x)$ are the eigenfunctions of $h_1(x_1)$ and $h_2(x_2)$ with eigenvalues $\varepsilon_n, \varepsilon_m,$ the eigenvalues $r_{n,m}$ of the symmetry operator $R^{(2)}$ are:
\be
r_{n,m}=(\varepsilon_{n}-\varepsilon_{m})^{2}+2(\varepsilon_{n}+\varepsilon_{m})+1;\quad \varepsilon_{n}=-(A-n)^{2}.\nonumber
\ee
After straightforward calculations,
\be
r_{n,m}=4((n-m)^{2}-1)(A-\frac{n+m+1}{2})(A-\frac{n+m-1}{2}).\label{r2}
\ee
It is clear that $r_{n,m}=0$ for $m=n\pm 1,$ i.e. the norm of $\Psi^{(1)}_{n, n\pm 1}$ vanishes. These wave functions of $H^{(2)}$ are simultaneously the zero modes of $Q^+,$ and the corresponding energy levels $E^{(1)}_{n, n\pm 1}$ (see (\ref{psi1})) are absent in the spectrum of $H^{(1)}.$ Also, one can make sure that the norm of other $(|n-m|>1)$ bound state wave functions with $n, m < A$ are positive: $r_{n,m}>0.$

Could some extra bound states exist besides those of (\ref{psi1}) ? If the Hamiltonian $H^{(1)}$ has such eigenstates, we have to take an interest in their superpartners among wave functions of $H^{(2)}.$ The superpartnership has to be provided by intertwining relations (\ref{int}).
Two possibilities are discussed below in the items (ii) and (iii).

(ii) It is easy to imagine the extra state of $H^{(1)}$ if the corresponding wave function turns out to be simultaneously the zero mode of operator $Q^-.$ Then, nothing unexpected: its possible superpartner wave function trivially vanishes. Therefore, we must look for the normalizable zero modes of $Q^-:$
\be
Q^{-}\Omega^{-}_{n}(\vec{x})=0. \label{omega}
\ee
This problem can be reduced to the problem with separation of variables by using the similarity transformation:
\ba
Q^{-}=e^{-\chi(\vec{x})}q^{-}e^{+\chi(\vec{x})};\qquad \Omega^{-}_{n}(\vec{x})=e^{-\chi(\vec{x})}\omega_{n}(\vec{x}),\label{sim2}
\ea
where
\be
\chi(\vec{x})=-\int C_{+}(x_+)dx_{+}-\int C_{-}(x_-)dx_{-}, \label{chi}
\ee
and $C_{\pm}(x_{\pm})$ were defined by (\ref{C}).

The variables in the supercharge are separated:
\be
q^{-}=\partial^{2}_{1}-\partial^{2}_{2}-U(x_{1})+U(x_{2}), \nonumber
\ee
reducing the two-dimensional problem to a couple of one-dimensional problems, but both with the same potential $U(x)$ given in (\ref{U}).
It was fortunate that these one-dimensional problems (with equal values of spectral parameters)
\ba
\biggl(-\partial^{2}_{1}+U(x_{1})\biggr) \rho^{(1)}_{n}(x_{1})&=&\epsilon_{n}\rho^{(1)}_{n}(x_{1});
\label{eta1}\\
\biggl(-\partial^2_2+U(x_2)\biggr) \rho^{(2)}_{n}(x_{2})&=&\epsilon_{n}\rho^{(2)}_{n}(x_{2})
\label{eta2}
\ea
coincide with equation (\ref{scarf}) whose solutions were already given by (\ref{eta}): $\rho^{(1)}_n(x_1)=\eta_n(x_1);\, \rho^{(2)}_n(x_2)=\eta_n(x_2)$.
As usual in the procedure of conventional separation of variables, the auxiliary zero mode $\omega$ of the operator $q^-$ is expressed as a linear combination of products $(\rho^{(1)}_n(x_1) \cdot \rho^{(2)}_n(x_2)).$ According to (\ref{sim2}), (\ref{C}) and (\ref{chi}), the desired zero modes $\Omega$ are:
\ba
\Omega^{-}_{n}(\vec{x})=|sech(x_{+}/2)| \cdot |cosech(x_{-}/2)| \eta_{n}(x_1)\eta_n(x_2). \nonumber
\ea
In contrast to the situation in Subsection 2.2, no antisymmetric combination can be used, and the singularity of $ cosech $ on the line $x_- = 0$
cannot be compensated. As result, all zero modes $\Omega_n$ are nonnormalizable, and the corresponding extra bound state wave functions from the option (ii) do not exist.

(iii) The extra bound state $\Psi^{(1)}$ of $H^{(1)}$ could also exist, if their superpartners $\Psi^{(2)}$ are nonnormalizable, i.e. if the operator $Q^-$ could transform normalizable wave function into nonnormalizable one. Since the singularity of $Q^-$ are concentrated on the line $x_- = 0,$
we have to consider the behaviour of operators and functions just in its neighborhood. In this vicinity, both Hamiltonians $H^{(1)},\, H^{(2)}$ are
amenable to separation of variables in terms of $x_{\pm}$:
\ba
H^{(1),(2)} &\sim & -2\bigl(\partial_{-}^{2}+\partial_{+}^{2}\bigr)+2\frac{a(a\mp 1)}{x_{-}^{2}}+V_{+}(x_{+}); \label{asymp1}\\
Q^{\pm} &\sim & \bigl(\partial_{+}\mp\frac{1}{2}\tanh(x_{+}/2)\bigr)\bigl(\partial_{-}\mp\frac{1}{x_{-}}\bigr), \label{asymp2}
\ea
where $V_+$ is nonsingular and depends on $x_+$ only. Due to separation (\ref{asymp1}), wave functions $\Psi^{(1), (2)}$ in the vicinity under discussion are represented as the sum of products: $\Psi^{(1),(2)} \sim \psi^{(1),(2)}_{-}(x_{-})\psi^{(1),(2)}_{+}(x_{+}).$ The explicit form of (\ref{asymp1})
means that nonsingular solutions $\psi^{(1)}_-(x_-)\sim x_{-}^{2}.$ It is clear that operator (\ref{asymp2}) is not able to destroy such normalizable behaviour of $\psi^{(1)}_-(x_-),$ i.e. the option (iii) for extra bound states of $H^{(1)}$ is not realized as well.

Summing up the results of the present Section, the discrete spectrum of $H^{(1)}$ with nonseparable variables is nondegenerate and consists of
 bound state levels with energies (\ref{psi1}) for $|n-m| \geq 2,$ and $n, m < A.$ The corresponding wave functions $\Psi^{(1)}_{n,m}=\Psi^{(1) +}_{n,m}$ (symmetrical under $x_1 \rightleftarrows x_2$) are given by (\ref{psi1}). This is the complete analytical solution for the discrete spectrum of quantum problem with Hamiltonian $H^{(1)}$ for $a=-1.$

\section{Exact solution of the models with $a_k=-k$}

The results of previous Section will be extended now for the whole hierarchy of Hamiltonians $H^{(1)}(a_k)$ with $a_k=-k,\, k=1,2,...,$ with the previous one $H^{(1)}\equiv H^{(1)}(a_1).$ Analogously, their superpartners wil be denoted as $H^{(2)}(a_k).$ Just the important relations between superpartners with different values of parameters - two-dimensional shape invariance \cite{gendenstein}, \cite{mallow}, \cite{new}, \cite{ioffe1}, \cite{shape2011} - allow to solve the quantum problems for $H^{(1)}(a_k).$ This property follows from the simple identity $a_{k}(a_{k}-1)=a_{k+1}(a_{k+1}+1)$ and it has the form:
\be
H^{(1)}(a_{k})=H^{(2)}(a_{k+1}). \label{shape}
\ee
Therefore, the infinite chain (hierarchy) of Hamiltonians can be built:
\be
H^{(2)}(a_{1})\div H^{(1)}(a_{1})=H^{(2)}(a_{2})\div H^{(1)}(a_{2})=...\div H^{(1)}(a_{N-1})=H^{(2)}(a_{N})\div H^{(1)}(a_{N})=...,\label{hierarchy}
\ee
where the sign $\div $ means that the corresponding Hamiltonians $H^{(1), (2)}(a_k)$ are intertwined by supercharges $Q^{\pm}(a_k).$
By means of combination of shape invariance (\ref{shape}) and intertwining relations (\ref{int}) recurrently moving along the chain (\ref{hierarchy}), we have to perform the full analysis analogous to that of Section 2.

(i) The first kind of wave functions $\Psi^{(1)}(\vec x; a_k)$ of the Hamiltonian $H^{(1)}(a_k)$ is obtained from known wave functions of $H^{(2)}(a_1)$ by the action of $k$ operators $Q^+$ with different values of parameters:
\be
\Psi^{(1)}(a_{k})=Q^{+}(a_{k})\Psi^{(2)}(a_{k})=...= Q^{+}(a_{k})Q^{+}(a_{k-1})...Q^{+}(a_{1})\Psi^{(2)}(a_{1}),\label{wave}
\ee
where $Q^+(a_k)$ are defined by (\ref{Q}) with $a=a_k;\,\, C_{+}(a_k)=\frac{1}{2}a_k\tanh(x_{+}/2);\quad C_{-}(a_k)=\frac{1}{2}a_k\coth(x_{-}/2).$
It is convenient to check the normalizability of functions (\ref{wave}) by calculating their norms.
The norm of wave function (\ref{wave}) is expressed as:
\ba
\|\Psi_{n,m}^{(1)}(a_{k})\|^{2}&=&
<\Psi_{n,m}^{(2)}(a_{k})|Q^{-}(a_{k})Q^{+}(a_{k})|\Psi_{n,m}^{(2)}(a_{k})> =\nonumber\\
&=&<\Psi_{n,m}^{(2)}(a_{1})|Q^{-}(a_{1})Q^{-}(a_{2})...Q^{-}(a_{k})Q^{+}(a_{k})...Q^{+}(a_{2})
Q^{+}(a_{1})|\Psi_{n,m}^{(2)}(a_{1})>= \nonumber\\
&=&<\Psi_{n,m}^{(2)}(a_{1})|\Lambda(a_{k})|\Psi_{n,m}^{(2)}(a_{1})>=\lambda_{n,m} \parallel\Psi^{(2)}_{n,m}(a_k)\parallel^2,\label{norm2}
\ea
where operator $\Lambda (a_k)$ is defined by the operator product in the second line of (\ref{norm2}), and for $k=1$ it coincides identically with the symmetry operator $R^{(2)}$ of the previous Section.
The proof of the last equality in (\ref{norm2}) and calculation of eigenvalues $\lambda_{n,m}$ are rather cumbersome, and this is described in Appendix. The result can be formulated as follows.
\ba
\lambda_{n,m}&=&0 \quad for \quad |n-m| \leq k \label{r11}\\
\lambda_{n,m}&>&0 \quad for \quad |n-m| > k. \label{r22}
\ea
This means that wave functions $\Psi^{(1)}_{n,m}(a_{k})$ of the first kind are normalizable for $|n-m| >k;\, n,m < A,$ and they vanish trivially for other $n,m.$ The corresponding nondegenerate energy eigenvalues are still given by (\ref{energy}).

(ii) For $a=a_k,$ the equation (\ref{omega}) for zero modes $\Omega_n^-(a_k)$ of operator $Q^-(a_k)$ allows again the conventional separation of variables due to similarity transformation by $\chi(\vec x; a_k)=-\int C_+(x_+; a_k)dx_+ - \int C_-(x_-; a_k)dx_-.$ Since the one-dimensional potential $U(x)$ in (\ref{U}) does not depend on parameter $a,$ the equations (\ref{eta1}), (\ref{eta2}) do not change, and explicit expressions for functions $\rho^{(i)}(x_i)$ coincide with (\ref{eta}). Then, the zero modes can be written as:
\be
\Omega^-_n(\vec x; a_k)= e^{-\chi(\vec x; a_k)}\omega_n(\vec x)= |\cosh(x_+/2)|^{a_k}|\sinh(x_-/2)|^{a_k}\omega_n(\vec x).\nonumber
\ee
It is evident, that for our case $a_k=-k$ these functions are singular and nonnormalizable. Again, no extra bound states of this kind exist for the Hamiltonian $H^{(1)}(a_k).$

(iii) The existence of this kind of bound states depends on opportunity that $Q^-(a_k)$ is able to destroy the normalizability of $\Psi^{(1)}(\vec x; a_k).$ It is necessary again to consider the behaviour of operators and wave functions around the line of singularity $x_- = 0.$ The asymptotic form of $H^{(1), (2)}$ and $Q^{\pm}$ is the same as in (\ref{asymp1}), (\ref{asymp2}), and the normalizable wave functions $\Psi^{(1)} \sim x_-^{k+1}$ in the vicinity of $x_- = 0.$ This behaviour can not be destroyed by the action of $Q^-(a_k).$ Thus, similarly to analysis in the very end of previous Section, this kind of extra wave functions also do not exist.

\section{Conclusions}

The two-dimensional quantum model, which can be called as two-dimensional Scarf II model, was demonstrated to be exactly solvable for arbitrary value
$a_k = -k.$ The spectrum of each member of this hierarchy of Hamiltonians is nondegenerate. The values of energies are given by (\ref{energy}) for $|n-m| > k,$ and the corresponding wave functions are given by (\ref{wave}). The complete analytical solution of this two-dimensional model, together with generalized Morse model \cite{physrev} and generalized P\"oschl-Teller model \cite{poschl2}, demonstrates that supersymmetrical approach is a powerful method to solve the problems which are not amenable to conventional separation of variables.

\section{Acknowledgements}

The work of M.V.I. and E.V.K. was partially supported by the grant RFFI 09-01-00145-a. E.V.K. is also indebted to the non-profit foundation "Dynasty" for financial support.

\section{\bf Appendix}

1. To calculate the matrix element (\ref{norm2}), we shall simplify the operator $\Lambda (a_k).$  Let us prove that:
\be
\Lambda(a_{k})=\Lambda(a_{1})\bigl(\Lambda(a_{1})+\Gamma_{2}\bigr)...\bigl(\Lambda(a_{1})+\Gamma_{2}...+\Gamma_{k}\bigr), \label{a1}
\ee
where $\Gamma_k$ is defined as:
\be
\Gamma_{k}=(2k-1)\biggl( 2H^{(2)}(a_{1})+2k(k-1)+1\biggr). \label{a2}
\ee
The explicit expressions (\ref{Q}) for supercharges $Q^{\pm}$ provide the following relations between $Q^{\pm}(a_{k})$ and $Q^{\pm}(a_{k-1}):$
\be
Q^{\mp}(a_{k})=Q^{\pm}(a_{k-1})\pm(2k-1)\biggl(2\tanh\frac{x_+}{2}\partial_- +2\coth\frac{x_{-}}{2}\partial_{+}\pm\tanh\frac{x_{+}}{2}\coth\frac{x_{-}}{2}\biggr)\nonumber
\ee
Multiplying these relations, we obtain:
\ba
& &Q^{-}(a_{k})Q^{+}(a_{k})=\nonumber\\
&=&Q^{+}(a_{k-1})Q^{-}(a_{k-1})-Q^{+}(a_{k-1})(2k-1)\biggl(2\tanh\frac{x_{+}}{2}\partial_{-}
+2\coth\frac{x_{-}}{2}\partial_{+}-\tanh\frac{x_{+}}{2}\coth\frac{x_{-}}{2}\biggr)+\nonumber\\
&+&(2k-1)\biggl(2\tanh\frac{x_{+}}{2}\partial_{-}
+2\coth\frac{x_{-}}{2}\partial_{+}+\tanh\frac{x_{+}}{2}\coth\frac{x_{-}}{2}\biggr)Q^{-}(a_{k-1})-\nonumber\\
&-&(2k-1)^{2}\biggl(2\tanh\frac{x_{+}}{2}\partial_{-}
+2\coth\frac{x_{-}}{2}\partial_{+}+\tanh\frac{x_{+}}{2}\coth\frac{x_{-}}{2}\biggr)\cdot\nonumber\\
&\cdot &\biggl(2\tanh\frac{x_{+}}{2}\partial_{-}
+2\coth\frac{x_{-}}{2}\partial_{+}-\tanh\frac{x_{+}}{2}\coth\frac{x_{-}}{2}\biggr).
\label{a4}
\ea
Definition (\ref{V}) of potential in $H^{(1)}(a_k)$ and simplification of the r.h.s. of (\ref{a4}) allow to rewrite it as follows:
\ba
& &Q^{-}(a_{k})Q^{+}(a_{k})=Q^{+}(a_{k-1})Q^{-}(a_{k-1})+(2k-1)\biggl(2H^{(1)}(a_{k-1})+2(k^{2}-k)+1\biggr)= \nonumber\\
& &=Q^{+}(a_{k-1})Q^{-}(a_{k-1})+\Sigma_k;\quad \Sigma_k\equiv (2k-1)\biggl(2H^{(1)}(a_{k-1})+2k(k-1)+1\biggr). \nonumber
\ea

The relation (\ref{a1}) can be proved now by mathematical induction. For $k=1$ and $k=2$ (\ref{a1}) is obviously satisfied. Let us prove
that it is true for $\Lambda (a_{k+1})$ under the assumption that it is fulfilled for $\Lambda(a_k).$ From the intertwining relations and shape invariance it follows that operators $\Sigma_k$ and $\Gamma_{k}$ are intertwined:
\be
\Sigma_k\cdot Q^{+}(a_{k-1})...Q^{+}(a_{1})=Q^{+}(a_{k-1})...Q^{+}(a_{1})\cdot\Gamma_{k}. \nonumber
\ee
This relation can be used in the following chain of transformations:
\ba
& &\Lambda(a_{k+1})=Q^{-}(a_{1})...Q^{-}(a_{k+1})Q^{+}(a_{k+1})...Q^{+}(a_{1})=\nonumber\\
&=&Q^{-}(a_{1})...Q^{-}(a_{k})\biggl(Q^{+}(a_{k})Q^{-}(a_{k})+\Sigma_{k+1}\biggr)Q^{+}(a_{k})...Q^{+}(a_{1})=\nonumber\\
&=&\Lambda(a_{k})\Gamma_{k+1}+Q^{-}(a_{1})...Q^{-}(a_{k})Q^{+}(a_{k})Q^{-}(a_{k})Q^{+}(a_{k})...Q^{+}(a_{1})=\nonumber\\
&=&\Lambda(a_{k})\Gamma_{k+1}+Q^{-}(a_{1})...Q^{-}(a_{k})Q^{+}(a_{k})\biggl(Q^{+}(a_{k-1})Q^{-}(a_{k-1})
+\Sigma_k\biggr)Q^{+}(a_{k-1})...Q^{+}(a_{1})=\nonumber\\
&=&\Lambda(a_{k})\biggl(\Gamma_{k+1}+\Gamma_{k}\biggr)+Q^{-}(a_{1})...Q^{-}(a_{k})Q^{+}(a_{k})Q^{+}(a_{k-1})\biggl(Q^{+}(a_{k-2})Q^{-}(a_{k-2})
+\Sigma_{k-1}\biggr)\cdot\nonumber\\
&\cdot &Q^{+}(a_{k-2})...Q^{+}(a_{1})= ... =
\Lambda(a_{k})\biggl(\Gamma_{k+1}+\Gamma_{k}+\Gamma_{k-1}\biggr)+Q^{-}(a_{1})...Q^{+}(a_{k})Q^{+}(a_{k-1})Q^{+}(a_{k-2})\cdot\nonumber\\
&\cdot &\biggl(Q^{+}(a_{k-3})Q^{-}(a_{k-3})+\Sigma_{k-2}\biggr)Q^{+}(a_{k-3})...Q^{+}(a_{1})= ... =
\Lambda(a_{k})\biggl(\Gamma_{k+1}+\Gamma_{k}+...+\Gamma_{3}\biggr)+\nonumber\\
&+&Q^{-}(a_{1})...Q^{+}(a_{3})Q^{+}(a_{2})\biggl(Q^{+}(a_{1})Q^{-}(a_{1})+\Sigma_2\biggr)Q^{+}(a_{1})=\nonumber\\
&=&\Lambda(a_{k})\biggl(\Gamma_{k+1}+\Gamma_{k}+...+\Gamma_{2}\biggr)+\Lambda(a_{k})\Lambda(a_{1})=
\Lambda(a_{k})\biggl(\Lambda(a_{1})+\Gamma_{2}+...\Gamma_{k+1}\biggr)=\nonumber\\
&=&\Lambda(a_{1})\biggl(\Lambda(a_{1})+\Gamma_{2}\biggr)...\biggl(\Lambda(a_{1})+\Gamma_{2}+...+\Gamma_{k+1}\biggr),\nonumber
\ea
which finishes the proof of relation (\ref{a1}).

2. Let us use the expression (\ref{a1}) for $\Lambda (a_k)$ and coincidence $\Lambda (a_1)=R^{(2)}(a_1)$ for calculation of eigenvalues $\lambda_{n,m}$ in (\ref{norm2}). The symmetry operator $R^{(2)}(a_1)$ can be replaced in (\ref{norm2}) by its eigenvalues $r_{n,m}$ from (\ref{r2}),
and sums of $\Gamma'$s:
\be
\Gamma_{2}+...\Gamma_{i}=\sum_{l=2}^{i}(2l-1)\biggl(2E_{n,m}+2l^{2}-2l+1\biggr)=(i^{2}-1)\biggl(2E_{n,m}+i^{2}+1\biggr), \nonumber
\ee
due to the definition (\ref{a2}) of $\Gamma_k.$

We are able now to evaluate the product for $n, m < A$:
\ba
& &r_{n,m}\prod_{i=2}^{k}\bigl(r_{n,m}+\Gamma_{2}+...\Gamma_{i}\bigr)=
r_{n,m}\prod_{i=2}^k \biggl(r_{n,m}+2(i^{2}-1)E_{n,m}+i^{4}-1\biggr)=\nonumber\\
&=&r_{n,m}\prod_{i=2}^k\biggl[4A^{2}\biggl((n-m)^{2}-i^{2}\biggr)+4A(n+m)\biggl((i^{2}-(n-m)^{2})\biggr)+\nonumber\\
&+&\biggl((n^{2}-m^{2})^{2}-2i^{2}(n^{2}+m^{2})+i^{4}\biggr)\biggr] =\nonumber\\
&=&r_{n,m}\prod_{i=2}^k \biggl[4\bigl((n-m)^{2}-i^{2}\bigr)\bigl(A-\frac{n+m+i}{2}\bigr)\bigl(A-\frac{n+m-i}{2}\bigr)\biggr]. \nonumber
\ea
All multipliers above are nonnegative but sometimes (for $|n-m|\leq k$) they vanish. Therefore, the norm (\ref{norm2}) of bound states $\Psi_{n,m}^{(1)}(a_k)$
with $|n-m|>k$ is positive, and they are normalizable, but the states with $|n-m|\leq k$ are absent in the spectrum of $H^{(1)}(a_k).$

\end{document}